\documentclass[a4paper,11pt]{article}
\usepackage{jheppub} 

\usepackage[utf8]{inputenc}
\usepackage{amsmath}
\usepackage{graphicx}
\usepackage{color}

\usepackage{amsmath,epsfig}
\usepackage{amssymb,amsfonts}
\usepackage{latexsym}

\usepackage{tocvsec2}
\usepackage{subeqnarray}
\usepackage{xcolor}

\usepackage{graphicx}
\usepackage{longtable}
\def\nn{\nonumber}

\relax

\usepackage[vcentermath,enableskew]{youngtab}
\def\hri#1#2{\href{http://arxiv.org/abs/#1}{[arXiv:#1 [#2]]}}
\def\hre#1#2{\href{http://arxiv.org/abs/#1/#2}{[arXiv:#1/#2]}}

\def\hrj#1#2{\href{www.doi.org/#1}{#2}}
\relax

\def\be{\begin{equation}}
\def\ee{\end{equation}}

\newcommand{\bear}{\begin{eqnarray}}
\newcommand{\bea}{\begin{eqnarray}}
\newcommand{\eear}{\end{eqnarray}}
\newcommand{\eea}{\end{eqnarray}}

\newbox\pippobox

\def\II{\relax{\rm I\kern-.18em I}}

\def\t{\tau}

\def\tr{\ensuremath{\mathrm{Tr}}}

\def\II{{\cal I}}

\title{\boldmath Wilson Loops and Wormholes}

\author[a]{Panos Betzios,}
\author[b]{Olga Papadoulaki}
\affiliation[a]{\href{https://phas.ubc.ca/}{Department of Physics and Astronomy} , University of British Columbia, \\
6224 Agricultural Road, Vancouver, B.C. V6T 1Z1, Canada}
\affiliation[b]{\href{https://perimeterinstitute.ca/}{Perimeter Institute for Theoretical Physics} , Waterloo, \\
 Ontario N2L 2Y5, Canada \\}
\emailAdd{pbetzios@phas.ubc.ca} \emailAdd{opapadoulaki@perimeterinstitute.ca}

\abstract{We analyse the properties of Wilson loop observables for holographic gauge theories, when the dual bulk geometries have a single and/or multiple boundaries (Euclidean spacetime wormholes). Such observables lead to a generalisation and refinement of the characterisation in \cite{Schlenker:2022dyo} based on the compressibility of cycles and the pinching limit of higher genus Riemann surfaces, since they carry information about the dynamics and phase structure of the dual gauge theory of an arbitrary dimensionality. Finally, we describe how backreacting correlated observables such as Wilson loops can lead to wormhole saddles in the dual gravitational path integral, by taking advantage of a representation theoretic entanglement structure proposed in \cite{BKP2,Betzios:2022oef}.}

\begin{document}
\maketitle
\flushbottom

\section{Introduction}
\label{sec:intro}

Euclidean ``spacetime" wormholes are fascinating objects presenting various difficulties in their physical interpretation\footnote{A small note on nomenclature and scope: The configurations we study are primarily defined in Euclidean signature and involve all the dimensions of the Euclidean spacetime. Nevertheless they 
can also contribute to Lorentzian quantities, since many field theoretic Lorentzian computations 
can be performed via the use of Euclidean techniques.}.
Even in the context of the AdS/CFT correspondence, for which we have an in-principle non-perturbative understanding of the bulk quantum gravity path integral, wormholes correspond to amplitudes computed on a connected bulk manifold, but whose conformal boundary contains several disjoint pieces. It is quite hard to interpret this peculiar result in holography, since on the one hand from a boundary perspective the boundary QFT's seem to be completely decoupled, but on the other hand a certain form of cross-coupling seems to exist due to the bulk connectivity\footnote{For Lorentzian wormholes, it is understood that the reason for the bulk connectivity is the entanglement between the otherwise decoupled QFTs. For an analogous statement in Euclidean signature, see \cite{BKP2},\cite{Betzios:2022oef} and section~\ref{BackreactedWLandworms} .}. This has led to various paradoxes the most famous being the so called factorisation paradox \cite{maldamaoz}.

One possible explanation on how one can understand the peculiar features of wormhole amplitudes from a dual QFT perspective is provided under the general umbrella of averaging, whether the average is an explicit one---usually defined for models of low dimensionality (see for example \cite{Betzios:2020nry,Maloney:2020nni,Chandra:2022bqq,Saad:2019lba,Cotler:2020ugk,Kundu:2021nwp})---or an effective one due to the chaotic nature of the dual QFT and the coarse graining that is implicit when using the low energy effective gravitational theory (see for example \cite{Schlenker:2022dyo,Belin:2020hea, Altland:2021rqn} and several references within). Most of these works concern computing very basic observables such as the partition function or averages of products of the partition function and their Laplace transform and the wormholes do not have to be saddle points of the low energy effective gravitational action. Another reason that the literature concerning observables and correlation functions of local and non-local operators on wormhole backgrounds has been meager so far, is precisely because of the difficulties in interpreting the results from a dual QFT perspective (see though \cite{BKP1, BKP2} for computations of more complicated observables on wormhole gravitational saddles).  

A possible way to distinguish the role of wormholes in computing various QFT observables was given in \cite{Schlenker:2022dyo} and posits that wormholes are relevant only for observables whose expectation value is affected by high energy/above the black hole threshold states in the dual QFT. This is practically imposed as a geometric criterion, by studying the compressibility properties of various ($S^1$) cycles of the boundary manifold and their behaviour under pinching. We review this criterion and the conclusions of \cite{Schlenker:2022dyo} in section~\ref{Contractpinch}. This geometric (pinching) criterion fails though if the first homotopy group of the boundary $\pi_1(\partial \mathcal{M})$ is trivial, such as when $\partial \mathcal{M} = S^d$ (or a disjoint union of such manifolds in the case of a wormhole geometry)\footnote{Even though in the work of \cite{Schlenker:2022dyo}, only $S^1$ cycles where considered, one can try to generalise this criterion to cycles of different dimensionality and higher homotopy groups, but the physical connection with black holes and the spectrum is harder to establish.}. In this work, we shall describe a more refined criterion based on the behaviour of line observables and in particular Wilson loops\footnote{As in the previous footnote perhaps the analysis of higher dimensional operators such as surface operators can give even further physical information about wormhole backgrounds.}. The advantage of the Wilson loop criterion is that it contains both the geometric (compressibility) information as well as more dynamical information about the dual QFT, through its numerical value and scaling properties. For example in the case of $S^d$ boundaries, when the pinching criterion is insufficient, the Wilson loop criterion becomes an indispensable tool that can distinguish the various phases and behaviours of the dual gauge theory. The various behaviours of Wilson (and Polyakov) loops for single and multi-boundary geometries are presented in sections~\ref{WLandHP} and~\ref{WLprobes}.

So far we assumed in our discussion that our line observables should be thought of as probes with which we can understand the properties of the dual QFT. In analogy with Lorentzian traversable wormholes \cite{Gao}, it is natural to expect that it should be possible to even construct Euclidean wormhole saddles when turning on appropriate interactions or considering correlated observables in otherwise decoupled copies of QFTs~\cite{BKP1,BKP2,VanRaamsdonk:2020tlr,VanRaamsdonk:2021qgv}. These correlated observables could take the form of local operator interactions smeared with a kernel $K(x-y)$~\cite{BKP1}
\be\label{correlatedlocal}
g \sum_i \int d^d x \int d^d y K(x-y) \mathcal{O}_i^{(1)}(x) \, \mathcal{O}_i^{(2)}(y) \, ,
\ee
where g is the coupling constant of the interaction and ${O}^{(I)}_{i}$ are operators of the $I$-th QFT copy.
A first difficulty with this construction, is that one needs in practice very ``heavy" sets of operators, so that their combined effect in \eqref{correlatedlocal} contributes to leading order in the $1/N$ expansion and results in a backreacted bulk saddle. In addition, if the kernel $K(x-y)$ is ultra-local $\sim \delta^d(x-y)$, the two boundaries are effectively brought in direct contact as in~\cite{Gao} and the holographic observables such as the two point cross-correlation functions typically cannot match the computations performed on Euclidean wormhole backgrounds (since in the latter there are no short distance singularities in the cross-correlators placed on distinct boundaries \cite{BKP1}). On the other hand non-local kernels or a smearing can work at an effective level, but we would also like to have a direct derivation of them from a progenitor UV complete model\footnote{In addition in some three-dimensional cases the backreacted wormhole geometries corresponding to heavy operator insertions involve the presence of conical defects in the bulk~\cite{Chandra:2022bqq}, so they are not strictly speaking completely smooth wormhole backgrounds.}. In~\cite{BKP2,VanRaamsdonk:2021qgv} some first steps towards resolving these issues were undertaken, by considering systems of holographic $BQFT_d$'s living on distinct $\Sigma_d$'s coupled via a ``messenger" $QFT_{d+1}$ living in one dimension higher (such as on $\Sigma_d \times I$), that do give rise to similar effective non-local cross-interactions upon integrating out the $QFT_{d+1}$, see~\cite{BKP2} for more details.  

Another interesting variation of this setup, that evades some of the issues mentioned above, is if one considers correlated observables of non-local operators such as Wilson loops or surface operators. In this case, what naturally replaces the label $i$ in eqn. \eqref{correlatedlocal}, are representations $R$ of the boundary QFT symmetry (gauge) group $\mathcal{G}$. One can then form ``entangled" sums of the general type~\cite{BKP2,Betzios:2022oef}
\be\label{correlatedloops}
Z_{system}^{(1-2)} \,= \,  \sum_R \, e^{-w(R)} \, \langle W_R^{(1)} \rangle_1 \, \langle W_R^{(2)} \rangle_2 \, ,
\ee
where $W_R^{(I)}$ is the non-local observable in the $(I)$-th copy of the $QFT_{d}$ and $w(R)$ is some weighting function that depends on the specific model. The label $R$ corresponding to representations of the dual field theory's symmetry group then plays the role of a ``superselection sector", similar to those that appear in the $\alpha$-state description of wormholes~\cite{Betzios:2022oef}. Such sums can either appear upon integrating out a ``messenger" quasi-topological $TQFT_{d+1}$ in one dimension higher (i.e. on $\Sigma_d \times I$), that couples the $QFT_{d}$'s (that are placed on distinct $\Sigma_d$'s) as in \cite{BKP2}, or it is also possible to view them as bona-fide insertions of correlated non-local observables in the path-integral of the otherwise decoupled $QFT_d$ copies\footnote{Another possibility is to consider a coherent superposition of such correlated sums and deform the action of the system with a term like~\eqref{correlatedloops} (now without the expectation values). By expanding the exponential one would find an infinite sum of terms with multiple (non-local) operator insertions in the path integral.}.

Of course, once more, we expect that these correlated non-local observables will lead to a backreacted geometry only when they are appropriately ``heavy" - but this can be easily achieved now, since there exist representations $R$ that contain $O(N^2)$ boxes in their Young diagram. A similar backreaction problem was actually analysed in great detail in the case of single boundary geometries dual to single half-BPS Wilson loop insertions in $\mathcal{N}=4$ SYM~\cite{DHoker:2007mci,Okuda:2008px,Aguilera-Damia:2017znn} as we review in section~\ref{BackreactedWL}. In section~\ref{BackreactedWLandworms} we briefly extend this discussion to the two boundary example of correlated Wilson loops between otherwise decoupled QFTs related to eqn.~\eqref{correlatedloops}\footnote{In upcoming work \cite{JiHoon} we analyse such examples in the context of $\mathcal{N}=4$ SYM in more detail both from the bulk and the dual field theory perspective.}, where a sum of such type appears.

\paragraph{Symmetries and moduli for wormhole backgrounds}

Let us now consider the general role of symmetries and their breaking on wormhole backgrounds, since they will play an important role in our analysis. These symmetries could be internal symmetries or isometries of the boundary manifolds. To start, a local Fefferman Graham expansion on a two-boundary wormhole would indicate two independently conserved stress energy tensors or currents:
$D_{\nu} T_1^{\nu \mu} = D_{\nu} T_2^{\nu \mu} = 0 \, , $ or $D_{\nu} J_1^{\nu} = D_{\nu} J_2^{\nu} = 0$, where $1,2$ label the two boundaries. On top of that the bulk diffeomorhism or Gauss' law constraints also need to be taken into account, giving rise to a relation between the integrals of the two stress energy tensors or currents on the two boundaries, for example
 $\int dx^{\mu} n^{\nu} (T_{1 \mu \nu} - T_{2 \mu \nu})  = 0 \, $.
 These relations constrain any holographic correlation functions. A similar story holds for any other global symmetries on the two wormhole boundaries.

We can now consider the product translation symmetry $\mathcal{T}_1 \times \mathcal{T}_2\, ,$ when the boundary theories are not coupled, or more generally the product $\textit{Isom}_1 \times \textit{Isom}_2$ with general boundary manifold isometries and $\mathcal{G}_1 \times \mathcal{G}_2$ the product of boundary global symmetries. The wormhole background breaks this product symmetry structure into its diagonal part i.e. $\textit{Isom}_{diag.} \, , \mathcal{G}_{diag.}$ and so forth. In the case of isometries, this diagonal part can be readily seen to be geometrised in the bulk for each radial slice and remains an unbroken symmetry of the background. 

Then there is the ``axial" symmetry part of relative isometry or general group transformations (also called ``twists"), for example let us call it $\textit{Isom}_{axial} \,$ or $\mathcal{G}_{axial}$.
This axial symmetry is broken by the wormhole background. If we view this as a spontaneously broken symmetry, then this means that the wormhole saddle should be viewed as a specific state of a theory that contains both connected (wormhole) and disconnected saddles (assuming that all the bulk fields respect the same asymptotic behaviour/boundary conditions for the two solutions). In such a case, due to the spontaneously broken axial symmetry, it is possible to find zero modes that need to be integrated over and hence there could exist a moduli space of solutions. If on the other hand for a specific model the axial symmetries are explicitly broken by interactions (and the wormhole saddle does not compete with the product of unfactorised ones) there are no such moduli. Of course for the unbroken diagonal symmetry there are no zero modes to integrate over.

Now in a construction involving direct interactions between the two boundary theories such as in eqn.\eqref{correlatedlocal} the isometry/translational moduli are lifted and there is only the $\text{Isom}_{diag.}$ remaining, giving rise to spacetime dependent two point cross correlators. The same is also true in wormhole models when we have indirect cross-interactions, such as in the ``slab" models of two holographic $BQFT_d$'s
coupled to a one dimensional higher ``messenger" $QFT_{d+1}$ on $\Sigma_d \times I$~\cite{BKP2,VanRaamsdonk:2020tlr,VanRaamsdonk:2021qgv} (assuming that the messenger theory is dynamical). In this case the two point cross correlators exhibit non-trivial momentum dependence and there are no twist/axial zero modes to be averaged over.

On the other hand, if there is a (twist/axial) moduli space of solutions, then this means that one has to average over them and they cannot have arisen from  an (in)-direct cross-interaction term in the field theory action (because it would have lifted them). In this case the cross correlators should be averaged over such moduli resulting for example to constant two point cross correlators. An interacting UV complete model in which this can happen, is the case that the higher dimensional ``messenger theory" is topological \cite{BKP2}, since in this example the two-point function is a sector/representation theoretic average over the product of one point functions which by boundary translational or $\textit{Isom}$ invariance are simply constants\footnote{We do not consider here models of in-principle averaged $CFT$'s. In such models a similar discussion about the presence of twist/axial moduli applies.}. This discussion clarifies that both options (presence or not of a moduli space of solutions to be integrated over) can be realised in different models.

The structure of our paper is as follows: We first review the role of various cycles and their compressibility properties in single boundary geometries in section~\ref{Singleboundary}, using various examples - the main being the Hawking-Page transition. Then we describe the more powerful Wilson loop criterion that provides us with further physical information about the phases of the dual gauge theory and close this section
with an analysis of Wilson loops in very large representations, when they backreact in the dual geometry. In section~\ref{multibworm}, we describe the compressibility criterion for multiboundary geometries and in section~\ref{WormholesandWL} we thoroughly analyse properties of various Wilson loop observables in the case of two boundary wormholes. When one considers appropriately correlated Wilson loop observables of large representations, then they are expected to backreact and form the wormhole geometry itself as described in section~\ref{BackreactedWLandworms}. Throughout our analysis of Wilson loops we pay special attention to whether the loop should be thought of as a spatial or Polyakov loop and on the related patterns of symmetry and its breaking.

\section{Curves and Wilson loops for geometries with a single boundary}\label{Singleboundary}

\subsection{The Hawking Page (HP) transition }\label{HPtransition}

In this section we first review the Hawking Page transition for manifolds $\mathcal{M}$ whose asymptotic boundary has the topology of $\partial \mathcal{M} = S^1 \times S^{d-1}$~\cite{Witten:1998zw} (see \cite{aharony} for more general cases). 

The thermal $AdS$ metric is
\be\label{thermalAdS}
ds^2_{AdS} = \left(1+ \frac{r^2}{L^2} \right) d \tau^2 + \frac{dr^2}{\left(1+ \frac{r^2}{L^2} \right)} + r^2 d \Omega^2_{d-1} \, , \qquad \Lambda = \frac{d(d-1)}{2 L^2} \, ,
\ee
and $\Lambda$ is the AdS cosmological constant.
We define the periodically identified Euclidean time $\t \sim \t + \beta$ (in this solution the period is arbitrary). $L$ governs the size of $AdS$. In the same coordinate system, the Euclidean $AdS$ black hole (BH) metric is
\bea\label{BHAdS}
ds^2_{BH} = \left(1+ \frac{r^2}{L^2} - \frac{c_d M}{r^{d-2}} \right) d \tau^2 + \frac{dr^2}{\left(1+ \frac{r^2}{L^2} - \frac{c_d M}{r^{d-2}} \right)} + r^2 d \Omega^2_{d-1} \, \nn \\
c_d = \frac{16 \pi G}{(d-1) Vol(S^{d-1})} \, . \qquad
\eea
In order to eliminate the conical singularity at the tip of the Euclidean black hole we find that the temperature is fixed as follows ($r_+$ is always the largest solution corresponding to the outer horizon of the BH)
\be\label{smoothtip}
\beta = \frac{4 \pi r_+ L^2}{d r_+^2 + (d-2) L^2} \, , \qquad  1+ \frac{r_+^2}{L^2} - \frac{c_d M}{r_+^{d-2}} = 0 \, . 
\ee
The (renormalised) on-shell action is proportional to the (renormalised) volume and one finds the (scheme independent) difference of the two on-shell actions to be
\be\label{Onshelldiff}
\delta S = S_{BH} - S_{AdS} = \frac{V(S^{d-1}) \left(L^2 r_+^{d-1} - r_+^{d+1} \right)}{4 G_N \left(d r_+^2 + (d-2)L^2 \right)}
\ee
This is positive for small $r_+$ and negative for large $r_+$ signalling that the leading saddle is that of AdS for large $\beta$ and the BH for small $\beta$. 

In the special case of $d=2$, this analysis was refined and extended in the recent work~\cite{Schlenker:2022dyo}, for asymptotic topologies of higher genus as well as wormhole manifolds with two or more disconnected asymptotic boundaries. We shall further review these possibilities in the next section. Focusing for now in the case where the conformal boundary is a torus $S^1\times S^1$, equations \eqref{smoothtip} and \eqref{Onshelldiff} simplify and we find
\be\label{Onshelldiffdtwodim}
\beta = \frac{2 \pi L^2}{ r_+ } \, , \quad \delta S = \frac{  \pi  \left( L^2 r_+ - r_+^3  \right)}{4 G_N   r_+^2  } = \frac{ \beta^2 - 4 \pi^2 L^2}{8 G_N \beta} = \frac{2 \pi L}{8 G_N} \left(\frac{\beta}{2 \pi L} - \frac{2 \pi L}{\beta} \right) \, .
\ee
In these formulae, we observe the duality exchange between the two cycles of the torus relating the (BTZ) BH and (thermal) AdS. Tuning $\beta$ we find that either one of them diverges to $- \infty$. In particular $S_{BH}$ diverges for small $\beta$ giving the dominant high energy saddle and $S_{AdS}$
diverges for large $\beta$ giving the dominant low energy saddle. This divergence was shown in~\cite{Schlenker:2022dyo}, to be a general feature that arises from the pinching limit of the corresponding compressible cycle on the manifold. Since these notions will play an important role in the rest of this work, we now turn to a more detailed definition of them, and study various examples through this lens.

\subsection{Compressibility of cycles and the pinching limit}\label{Contractpinch}

Since our main object of interest will be Wilson loops, in this section we focus first on the information that we can glean just by geometric considerations related to closed curves on the asymptotic boundary $\partial \mathcal{M}$ and their pinching limit~\cite{Schlenker:2022dyo} and then move forward to analyse the additional dynamical information contained in the Wilson loops.

We define an arbitrary boundary closed curve
(a loop) $\gamma$ with boundary length $\ell_\gamma$.
Sometimes such a curve can wrap a non-trivial cycle $\mathcal{C}$ of the boundary manifold $\partial \mathcal{M}$ and in such cases it cannot be contracted to zero size (if the curve remains on the boundary). These cases exist when the boundary has a non-trivial fundamental (homotopy) group
$\pi_1(\partial \mathcal{M})$\footnote{Other homotopy groups $\pi_n(\partial \mathcal{M})$ would become relevant, if we were considering higher dimensional operators on the boundary, such as surface operators etc.}. Of course there always exist curves that do not wrap any non-trivial boundary cycle and are therefore boundary contractible.

When the loop wraps a non-trivial cycle $\mathcal{C}$, we define its pinching limit, as the one for which its physical size
(size of the loop) is driven to zero i.e. $\ell_\gamma \rightarrow 0$. For a boundary with the topology of an arbitrary compact Riemann surface of genus $g$ ($\partial \mathcal{M}  = \Sigma_g $), boundary pinching, either splits the surface in disconnected components or reduces the genus of that Riemann surface. In higher dimensions $d>2$, in many physically relevant examples the boundary manifold has at most a single non-trivial cycle: the thermal cycle $\mathcal{C}_\beta$\footnote{In other cases, such as when $\partial \mathcal{M} = S^d$, or a set of disconnected hyperspheres, there is no-non trivial cycle. This is one example that motivates us to perform a study of Wilson loops on the boundary manifold to characterise the various saddles/phases, since they contain more (dynamical) information about the boundary QFT.}. 

A boundary loop/curve $\gamma$ wrapping a cycle $\mathcal{C}$ is said to be compressible in the bulk - or bulk contractible, if we can shrink/contract it to zero size (a point) by deforming it continuously into the bulk (keeping the bulk manifold $\mathcal{M}$ fixed). In other words, even though from a purely boundary perspective such cycle is non-trivial, it in fact does not carry any non-trivial winding number, when the curve is allowed to contract into the bulk. 

Pinching a boundary cycle (reducing it to zero size), sometimes (but not always) leads to pinching the corresponding bulk geometry. For example let us consider a handlebody\footnote{These are (Schottky) manifolds $\mathcal{M}$ that when embedded in $\mathbb{R}^3$, they are topologically equivalent to the interior of their boundary $\partial{\mathcal{M}}$.} type of geometry (such as a solid torus). In such an example, when we pinch a compressible cycle $\mathcal{C}$, we inevitably pinch the bulk geometry. This can have as an effect of either separating the bulk manifold into disconnected pieces or simply changing the topology of the manifold  (reducing its genus). For an example of the first case, we can consider a handlebody $\mathcal{M}$, whose boundary is a genus two Riemann surface and we pinch the compressible cycle that splits both the boundary and the bulk manifold into two separate pieces (two torus handlebodies).
An example of the first type is pinching one torus handle of the genus two handlebody, giving rise to a new handlebody whose boundary is now one genus less - i.e. a torus.

On the other hand non-compressible cycles cannot be pinched in the same fashion. For example, if our cycle is a thermal circle that is not compressible in the bulk (thermal AdS), the pinching limit would simply result in a dimensional reduction of the bulk manifold $\mathcal{M}$. One can nevertheless argue physically, that as the non-compressible cycle becomes smaller and smaller, after a point there is another saddle that takes over and dominates in the path integral/thermodynamic ensemble (black hole). This later saddle allows taking a consistent pinching limit, since the corresponding cycle has now become compressible. We expect this statement to be true even for higher genera and/or multiple boundaries, since manifolds for which the cycle at question is compressible in the bulk, have infinite renormalised volume in the pinching limit of that cycle, while manifolds for which the boundary cycle to be pinched is non-compressible in the bulk, continue to have a finite renormalised volume as the boundary cycle is driven to zero size, as shown in~\cite{Schlenker:2022dyo} (see appendix~\ref{Pinchinggenustwo} for more details on the renormalised volume and its relation to the on-shell action). This means that the on-shell action of the former will always dominate the on-shell action of the later in the pinching limit and hence these are the saddles that play the dominant role in this limit.

If one gives an interpretation of the boundary cycle $\mathcal{C}$ as being a spatial cycle, this means that if we consider slicing our manifold on that cycle, we can define a state of the dual CFT. It was further argued in~\cite{Schlenker:2022dyo} that if the cycle is compressible in the bulk, this means that the state propagating through that cycle is a low energy (sub black hole threshold) fixed energy state. This is easily understood from the fact that the on-shell action diverges in the pinching limit of that cycle. The same divergence also appears in the dual CFT, since in this limit (the region of $\partial \mathcal{M}$ near this pinched cycle $\mathcal{C}$ resembles that of an an infinitely thin cylinder), it is the identity operator that flows through that cycle and the dual partition function exhibits the same kind of divergence. On the other hand non-compressible (spatial) cycles
can support above threshold high energy states flowing through them. A basic example for a Riemann surface of genus two and its pinching limit is analysed in more detail in appendix~\ref{Pinchinggenustwo}.

\subsection{Wilson Loops and the HP transition}\label{WLandHP}

In this section we shall assume that the boundary gauge theory in question has an $SU(N)$ (or $U(N)$) gauge symmetry. In this case there is another order parameter that we can use to distinguish the two phases of the Hawking-Page (HP) transition, the expectation value of a Wilson loop \cite{Witten:1998zw,Aharony:1998qu}. This expectation value in holography is computed by considering the (regularised) area of a string worldsheet that ends on the prescribed loop at the boundary, which we express with the following compressed formula
\be\label{WLexpect}
\langle W(\mathcal{C}) \rangle = \int_{worldsheets} d \mu \, e^{- A(D)} \, , \quad \partial D = \mathcal{C} \, ,
\ee
where $d \mu$ is the measure in the space of worldsheets and $A(D)$ is the Nambu-Goto action
for worldsheets of disk topology $D$ that end on the prescribed contour/cycle $\mathcal{C}$ on the boundary.
It is easy to observe that if the cycle that is pinched is not compressible in the bulk, the Wilson loop that is wound around that cycle, has zero expectation value, since there does not exist a minimal world-sheet bulk surface $D$ that ends on such a loop. On the other hand for a compressible cycle we can always find a minimal bulk worldsheet surface that ends on the boundary loop giving a priori a finite contribution after {renormalization}. We therefore find that another object that can distinguish compressible to non-compressible cycles is the Wilson loop and its expectation value serves as an order parameter. The existence of a bulk worldsheet surface ending on the Wilson loop is equivalent to the question of the compressibility of the corresponding cycle and can therefore be used as an equivalent criterion. 

Since the Wilson loop contains further dynamical information about the dual gauge theory, at this point one should distinguish whether the loop is a spatial or a temporal/Polyakov loop.

\paragraph{The thermal/Polyakov loop and center symmetry} The expectation value of a loop wound around the thermal circle ($\mathcal{C} = P \times S^1_\beta \, ,$ with $P$ a point on the transverse space) is related to the pattern of center symmetry breaking for the dual gauge theory. In particular a non zero expectation value for the thermal/Polyakov loop signals a center symmetry breaking, since a center element $g_c \in Z_N$\footnote{This can be represented (for $SU(N)$) as $g_c = z_i I_{N \times N} \, ,$ with $z_i^N = 1$.} transforms the Wilson loop as $\langle W_P(\mathcal{C} )\rangle \rightarrow g_c \langle W_P(\mathcal{C}) \rangle $\footnote{For a loop transforming in a representation $R$ with n-ality $n$, the center symmetry transformation acts as $\langle W_R(\mathcal{C} )\rangle \rightarrow g_c^{|R|} \langle W_R(\mathcal{C}) \rangle $, with $|R|$ the number of boxes in the rep and the n-ality given by $n= |R| \, mod \, N$. }. A non zero expectation value also means that the corresponding quark configuration has finite energy $\langle W_P(\mathcal{C} ) \rangle \sim e^{- \beta F_q}$ and thus the fundamental charges are not confined and exist and propagate in the spectrum of the theory. This typically happens in the high temperature phase of a gauge theory. On the other hand a zero expectation value for such a loop is indicative of a confining behaviour for the gauge theory. For the center symmetry to break one needs to take a certain thermodynamic limit, either that of an infinite volume or large-N. Let us then consider a gauge theory such as the $\mathcal{N}=4$ SYM on a product manifold $S^1_{\beta} \times \mathcal{M}_s$, with the spatial manifold $\mathcal{M}_s$ compact. It is then possible (at weak coupling) to integrate out all the matter fields to derive an effective action for the zero modes of the gauge field $A_{\tau}$ around the thermal circle~\cite{kp1,kp2} 
\be
 P e^{i \oint d \tau A_\tau} =  U \, . 
\ee
In this case, the partition function reduces to a unitary integral
\be
Z_{system} = \int D U e^{- S_{eff.}(U)} \, ,
\ee
with eigenvalues on the unit circle, that can in general be analysed at large-N with matrix model techniques. The resolvent and the density of eigenvalues on the unit circle $\rho(\theta)$ dictate the various phases and physics of the model. In particular an unbroken center symmetry means that
\be
\rho(\theta) = \frac{1}{2 \pi} \quad \Rightarrow \quad \frac{1}{N} \langle \tr U \rangle \vert_{N \rightarrow \infty} = \int_{-\pi}^\pi d \theta \rho(\theta) e^{i \theta} = 0 \, , 
\ee
so that center symmetry translates into the $U(1)$ translational symmetry for the density of eigenvalues at large-N. In the deconfined phase on the other hand, one finds that the support of $\rho(\theta)$ is gapped and $\langle \tr U \rangle \neq 0$ is a non-trivial function.

Performing the holographic dual bulk calculation to access strong coupling, one finds that for thermal AdS this expectation value is zero, while for the BH a naive calculation indicates a non-zero result. This is because in the case of AdS the thermal $S^1_\beta$ is not contractible in the bulk and hence there is no bulk worldsheet disk $D$ whose boundary is $S^1_\beta$. The situation is reversed for the BH, since the cigar geometry has the topology of a disk and one can thus readily find such a bulk worldsheet that covers this disk.

On the other hand we know that as long as the volume of the transverse spatial manifold (such as $S^{d-1}$) is finite, center symmetry cannot be spontaneously broken in normal circumstances due to the Gauss' law constraint\footnote{For a gauge theory on $S^3$ at weak coupling, one can indeed find a phase transition between a center symmetric and non-center symmetric distribution of eigenvalues around the unit circle for the holonomy of the Wilson loop \cite{kp1,kp2}. This result can be interpreted as a form of screening that happens due to taking the thermodynamic limit of large-N, effectively relaxing the gauge constraint and allowing the presence of a deconfined phase.}. A more refined computation shows that the expectation value of the Wilson loop indeed vanishes in this case due to an average over the zero mode of the $B$ field that couples to the bulk worldsheet~\cite{Witten:1998zw}. The particular coupling is of the form
\be
e^{i \int_D B} = e^{i \Phi} \, ,
\ee
and one has to average over the phase (zero mode) $\Phi$, resulting to a zero result for the expectation value of the loop\footnote{There is a subtlety here, in many cases (such as in $\mathcal{N}=4$ SYM) the center of the gauge theory is $Z_N$ and not $U(1)$ as this argument naively suggests. The resolution of this problem was given in \cite{Aharony:1998qu}, the argument being that certain objects act as a baryon vertex~\cite{Gross:1998gk} (for example a wrapped ($NS5$) fivebrane). In these cases the center remains $Z_N$ and there exist certain worldsheets that can end on this vertex (when their number is a multiple of $N$).}.

In the infinite volume limit of the $S^{d-1}$, one does recover a non-zero expectation value for the Polyakov loop, since there is no such zero mode for the $B$ field to average over (it becomes non-normalisable) and its value is fixed by asymptotic boundary conditions that are frozen. Nevertheless even at finite volume, we can still distinguish these two cases in the sense that when the cycle is compressible we can find a dual bulk minimal surface, while if it is not then such a surface does not even exist to start with, making clear the difference between the two phases.

\paragraph{Spatial Wilson loops}
The other type of loops we can consider are spatial loops, which exist for any Euclidean boundary manifold. In this case a boundary area law $\langle W(\mathcal{C}) \rangle \sim e^{- \sigma A(D_\mathcal{C}^{bound.})} $ would be indicative of confining behaviour, while a perimeter law $\langle W(\mathcal{C}) \rangle \sim e^{- p \ell(\mathcal{C})} $ of deconfined behaviour, but there can exist various other more complicated behaviours. In particular, for AdS at low temperatures one finds a Coulomb force between quarks, the simple argument being that even though the bulk configuration always scales as the surface of the disk shaped worldsheet $A(D)$,
one can use the background conformal invariance to rescale the worldsheet surface $D$ without changing its boundary (regularised) area $A(D_\mathcal{C}^{bound.})$. This stops being true in the case that the bulk geometry has an additional IR scale and gets effectively cutoff, or bounces (as in wormhole geometries), at this IR scale $\Lambda_{IR}$. In these examples, one indeed finds an area law indicative of a confining behaviour~\cite{BKP1}.

The case of $d=2$ is special, since one can exchange the meaning/nature of a wound loop along a cycle $\mathcal{C}$ as being either temporal or spatial\footnote{This holds also for the special case of any $d$-dimensional torus.}. For example in the case of a torus either cycle can play the role of Euclidean (compactified) time. 

We conclude this section by summarising that the Wilson loop observables contain both the geometric information regarding the compressibility of the corresponding cycle (at least at strong coupling where the geometry is trustworthy), as well as more refined dynamical information of the dual field theory that is encapsulated in its expectation value and scaling properties.

\subsection{Geometries dual to ``heavy'' backreacted Wilson Loops}\label{BackreactedWL}

In this section, we discuss a different limit for Wilson loops, in which they do backreact strongly on the background geometry. Such ``heavy'' Wilson loops are traces/characters of general irreducible representations $R$ of the boundary gauge symmetry group. The relevant irreducible representations for backreacting Wilson loops contain an $O(N^2)$ number of boxes in their Young diagrams with $O(N)$ boxes in each row/column. 

\paragraph{Single Wilson loop on $S^4$}

It is possible to analyse in detail such heavy Wilson loops, when the field theory at question has enough supersymmetry. The simplest case is the case of $\mathcal{N}=4$ SYM, where the expectation value of a single $1/2$-BPS Wilson loop placed at the equator of $S^4$ and transforming in the representation $R$ is computed in terms of the following hermitian matrix integral after performing supersymmetric localization \cite{Pestun}
\be\label{halfBPS}
\langle W_R \rangle = \frac{1}{Z} \int D M  e^{- \frac{2 N}{\lambda} \tr M^2} \tr_R \left( e^{M} \right) \, .
\ee
In particular we are interested in Wilson loops with very large representations with $O(N^2)$ boxes (both the rows and columns of the Young diagram describing the representation $R$ carry $O(N)$ boxes).

The holographic dual (backreacted) geometries of type-IIB string theory were constructed and classified in~\cite{DHoker:2007mci}. These are certain kind of ``bubbling'' geometries that preserve a bosonic $SO(2,1) \times SO(3) \times SO(4)$ subgroup of the bosonic part of the superconformal group $SO(2,4) \times SO(6)$ (the same symmetry that is preserved by the $1/2$-BPS Wilson loop in $\mathcal{N}=4$ SYM), and contain various non trivial fluxes and cycles. The metric of these backgrounds takes the form
\be\label{metricansatz}
    ds^2 = f_1^2 ds_{AdS_2}^2 + f_2^2 ds_{S^2}^2 + f_4^2 ds_{S^4}^2 + ds_{\Sigma}^2,
\ee
where the $AdS_2$, $S^2$, and $S^4$ are nontrivially fibered over the Riemann surface $\Sigma$ with coordinates $z, \bar{z}$ (that is usually taken to be the lower half-plane). This means that all the functions
depend on the Riemann surface coordinates $z,\bar{z}$ i.e. $f_i(z,\bar{z})$. In addition all the metric functions and fluxes are uniquely determined in terms of just two harmonic functions $h_{1,2}(z,\bar{z})$ on $\Sigma$, via a set of relations that can be found in~\cite{DHoker:2007mci}.

If one wishes to describe a background dual to a single backreacted loop, the $AdS_2$ factor then has the topology of a disk, whose boundary $S^1$ is the locus where the loop resides. A simple pole in the holomorphic part of $h_1$ corresponds to an asymptotic $AdS_5 \times S^5$ region of the metric \eqref{metricansatz}. In particular one combination of the coordinates in $\Sigma$ plays the role of a radial coordinate and together with the $AdS_2$ and $S^2$ factor form the $AdS_5$ portion of the geometry, while the other combination plays the role of an angular coordinate which together with the $S^4$ make up an $S^5$. By studying this asymptotic region, one finds that the loop is placed on a great $S^1$ of the asymptotic $S^4$ where the $\mathcal{N}=4$ super Yang-Mills resides.

In this example it is also possible to study in great detail how the backreacted Wilson loop observable directly affects the topological and geometric properties of the backreacted bulk manifold. What happens is that the planar resolvent $\omega_R(z)$ and spectral curve $y_R(z)$ of the matrix model \eqref{halfBPS} depend on the shape of the Young diagram of the representation $R$
and at the same time encapsulate all the data of the dual bulk geometry~\cite{Okuda:2008px,Aguilera-Damia:2017znn}, being identified with the holomorpic part of the harmonic functions $h_{1,2}$ via the relations
\be
2 \omega_R(z) = V_{class.}' - y_R(z) = \frac{2}{\lambda} z -   y_R(z) \, , \quad i y_R(z) =  h_1(z) \, , \quad \frac{2 i}{\lambda} z = h_2(z) \, .  
\ee
This means that there is a one to one map between shapes of Young diagrams and backreacted dual geometries. See~\cite{Aguilera-Damia:2017znn} for a detailed review containing various computations that can be performed (and matched) on both sides of this duality.

In the next, we discuss briefly two interesting extensions of this connection, more details will be provided in~\cite{JiHoon}.

\paragraph{$AdS_2$ with cylindrical topology and pairs of Polyakov loops} As a first extension of the existing results in the literature, let us now consider the $AdS_2$ factor of the metric ansatz \eqref{metricansatz} to have the topology of a Euclidean cylinder $I \times S^1$ and the metric
\be
ds^2_{AdS_2} = \frac{1}{\cos^2 \theta} \left(d\tau^2 + d \theta^2 \right) \, , \qquad - \pi/2 < \theta < \pi/2 \, , \quad \tau \sim \tau + \beta  \, . 
\ee
In this case the asymptotic form (we reach the asymptotic boundary by sending the coordinate $x \rightarrow \infty$ of $\Sigma_{z, \bar{z}}$ parametrised by $z=x+i y$) of the ten dimensional metric~\eqref{metricansatz} can be written as
 \be
ds^2 \sim  \left(d x^2 + \, \frac{e^{2 x}}{\cos^2 \theta}\left( d \theta^2 + d \tau^2 + \cos^2 \theta (d \phi^2 + \sin^2 \phi d \chi^2) \right)+  \, ds^2_{S^5}  \right) \, ,
\ee
with $\theta \in [-\pi/2, \pi/2]$. The topology of the large $x$ slices is that of $S^1 \times S^3$ and we find two additional boundary components at $\theta = \pm \pi/2$ (irrespective of $x$) that are connected through the $AdS_2$ bulk. This realises a geometry that is similar to the throat of \cite{Maldacena:2018gjk}. In contrast with that work though, the two $AdS_2$ boundaries are now part of the true asymptotic boundary and not cutoff and glued on a single common base space. The field theory dual is $\mathcal{N}=4$ SYM on $S^3 \times S^1_\beta$. An important point is that due to the Gauss' law on the compact $S^3$, the total charge carried by the loop operators (that appear as point particles on $S^3$) must vanish. In particular, the simplest non-trivial insertion is that of a pair of Polyakov loop operators wound around $S^1_\beta$ and carrying opposite charges (conjugate representations) one at the north and the other at the south pole of $S^3$. When they backreact on the geometry, these two loop operators lead to the presence of the two $AdS_2$ boundaries and the points $\theta = \pm \pi/2$ simply correspond to the north and south pole of the asymptotic $S^3$, where the two loops reside. This is the field theoretic reason why in this case one finds two $S^1$ boundaries in the $AdS_2$ factor of the geometry.

\paragraph{Multiple (non-interacting) clones of $\mathcal{N}=4$ SYM}
It is possible to extend this construction by considering multiple copies of $\mathcal{N}=4$ and considering heavy loop observables that correlate them in a representation theoretic fashion, see section~\ref{WormholesandWL}. In the case of $S^1_\beta \times S^3$ topology, the field theory dual should be described by an appropriate index containing multiple (non cross-interacting) copies of $\mathcal{N}=4$ with the insertion of Polyakov loops (around the $S^1_\beta$) that are correlated between the various copies. Is is interesting to study this technically tractable setup further both from the field theory as well as the bulk side.


\section{Multi-boundary geometries (wormholes)}\label{multibworm}

\subsection{Various cycles and the competition between factorised and non-factorised geometries}

Moving now to the case of wormholes with two boundaries, it was argued in~\cite{Schlenker:2022dyo}, that they do not contribute to the expectation value of observables which are not affected by above-threshold black hole states. In particular based on our analysis in section~\ref{Contractpinch}, the infinite renormalised volume in the pinching limit, or equivalently the compressibility of the cycle in the bulk reveals the presence of a sub-threshold state propagating (as is the case of a state in Euclidean AdS for which the cycle dual to the thermal circle defines a low energy state---a compressible cycle). On the other hand, the cycle dual to the thermal cycle in the BTZ case is non-compressible and defines a high energy state. Euclidean spacetime wormhole geometries with two boundaries have various non-compressible cycles
and hence by that argument are related to states that are above the black hole threshold (all the sub-threshold states correspond to dual geometries with compressible spatial cycles). A simple example to understand why Euclidean spacetime wormholes typically have non compressible cycles is the following: both the BTZ black hole and AdS can be viewed as a solid torus (filled donut) in $R^3$ and contain a single contractible cycle, so one can propagate either low energy or high energy states depending on which one is chosen to be the spatial cycle. On the other hand the topology of a $T^2 \times I$ two boundary wormhole can be obtained, by considering a solid torus with a smaller solid torus excised near its center. This creates another $T^2$ boundary in the interior. It is easy to see then that if we take a boundary cycle $\mathcal{C}_1$ on the first exterior boundary, it is not possible to retract it to zero size by moving in the interior of the bulk manifold and we just reach another cycle on the other torus boundary. Such a geometry has finite renormalised volume in the pinching limit\footnote{The same argument also applies to $T^2 \times I$ wormholes where both boundaries are large and of the same size.
An issue with the $T^2$ boundary wormholes (but not with the higher genus ones), is that they are not saddles of pure gravity alone, and one needs to add some additional background field to support them, or introduce appropriate defects in the bulk. Given such a construction, one would have to add counterterms and perform holographic renormalisation on both boundaries in order to compute their renormalised on-shell action and volume.} and does not describe sub-threshold states according to the argument above. 

There is an analogous statement, from a slightly different perspective that emphasizes the competition between factorised and non-factorised geometries (wormholes). If there is a system admitting a Euclidean wormhole saddle, we expect this saddle to possibly compete with a product of factorised saddles (if we can consistently impose the same set of boundary conditions for the fields on the corresponding disconnected boundaries for both types of solutions). We shall call the last type of single boundary saddles ``\emph{IR capped saddles}'', since they smoothly close/cap off in the IR.

A special case of an ``IR capped saddle'' is that of a black hole (that contains a thermal cycle $S^1_\beta$), or its double analytic continuation---the AdS soliton~\cite{Witten:1998zw}---but more generally one could consider other boundary topologies on which the dual QFTs live (such as simple spheres $S^d$), for which the bulk theory contains some other type of IR capped solution\footnote{Typically these would be related to a confining gauge theory with an IR mass gap/scale.}. The corresponding statement then is that spacetime wormholes will compete with such factorised but capped in the IR saddles. This is also in agreement with \cite{BKP2} that shows that Euclidean spacetime wormholes can be found in systems of confining/gapped QFT's, see for more details section~\ref{WormholesandWL}. The dissociation of the connected wormhole bulk geometry then has the interpretation of a dissociation at high energies of an effectively cross coupled system that decouples into two non cross-interacting sub-systems. The finite temperature case is simply a special subcase for which the conformal symmetry is broken due to the temperature (or in the double analytic continuation of the AdS Soliton~\cite{Witten:1998zw} by boundary conditions that drive confinement). This also leads us to conjecture that one has to inevitably break conformality either spontaneously or explicitly in $d \geq 2$ holographic (non-averaged) QFT systems, so that their dual bulk gravitational description can contain Euclidean wormhole saddles.

\section{Wilson loops and wormholes}\label{WormholesandWL}

\subsection{Wilson Loops as probes of a Euclidean wormhole geometry}\label{WLprobes}

We shall now uncover some further physical properties of Euclidean wormhole geometries by analysing correlation functions of boundary Wilson loops.

\paragraph{The case with no $S^1$ factor on the boundaries} In the simplest examples of Euclidean wormhole geometries, when the boundaries do not contain an $S^1$ factor and the first homotopy group is trivial i.e. when they have a spherical topology $S^d$, there is only a single notion of (spatial) Euclidean loop\footnote{One typically needs additional fields in the bulk such as axions or magnetic fluxes to support such geometries.}. In~\cite{BKP1} it was observed that the expectation value of such a loop obeys an area law when the loop size becomes large\footnote{In the absence of any other (boundary) scale the relative loop size is determined with respect to the size of the $S^d$.}, indicative of a confining behaviour.
 In this case there is also an IR (bulk) scale set by the minimum size of the bulk manifold at the point where the scale factor bounces (wormhole throat). In the parameter limit where the bulk manifold pinches off and becomes that of two decoupled AdS spaces, this area law ceases to exist. We should also emphasize once more that the compressibility criterion of \cite{Schlenker:2022dyo} does not apply at all in this case, since there are no non-trivial cycles on the boundary manifold.

 On Euclidean wormhole backgrounds there is a new additional possibility that one should analyse which provides us with further physical information, that of a two loop average $\langle W (\mathcal{C}_1)\, W (\mathcal{C}_2) \rangle$, each loop placed on a distinct boundary\footnote{One could also place the two Wilson loops on the same boundary. This case is quite similar to that when having a single boundary: for small Wilson loops at small distance there may be a single surface connecting them, while at larger distances there typically exist two distinct surfaces. In the wormhole example, if the loops are not wound around non-trivial boundary cycles the analysis is then similar to \cite{Drukker:1999zq,Gross:1998gk}. If one of the loops is wound around a non-trivial cycle and the other has trivial homology no connected surface exists. There can exist more general possibilities that are more complicated to visualise.} \cite{BKP1}. In the $S^d$-sliced wormhole  case, it was observed that there is a phase transition between a factorised set of two minimal surfaces each with a disk topology and one minimal surface with the topology of a cylinder. The second connected configuration dominates for large loops that probe the IR structure of the Euclidean QFT dual to the wormhole. This behaviour was termed \emph{a cross confining behaviour}, since this configuration corresponds to heavy quark/anti-quark pairs placed on the two distinct boundaries. In the UV (when the circular trajectories/loop sizes are small) these are independent and interact very weakly (in the bulk this manifests as a perturbative supergraviton exchange between two disconnected surfaces). In the IR (when the circular trajectories become large) their interaction becomes strong and the pair becomes bound, which manifests in the bulk by the development of a single world-sheet surface connecting the pair of loops. If we were to replace the two circular loops with two rectangular ones, this bound state could have a possible interpretation of a bound ``tetraquark" formed out of quark/anti-quark pairs of the two gauge theories on the two boundaries.

\paragraph{Boundaries with $S^1$ factors}
When the boundary/ies contain non-trivial $S^1$ factors, the analysis becomes more interesting, since one can also discuss Polyakov loops wound around the (thermal) cycles. The simplest example to analyze that still retains the complexity of this possibility is in $d=2$, for boundary slices having the topology of a (compact) Riemann surface of genus $g$. In contrast with the single boundary manifolds that we encountered in section~\ref{Contractpinch}, simple two boundary Euclidean wormhole geometries with a bulk $Z_2$ reflection symmetry do not have contractible cycles in the bulk, and one can argue that there is no corresponding worldsheet surface $D$ ending on a boundary loop that wounds around such a cycle $\mathcal{C}$. This is true for the one point function $\langle W(\mathcal{C}) \rangle$\footnote{Of course we could also consider loops that are not wound around the cycles $\mathcal{C}$ of the Riemann surface, and for these loops a minimal surface does exist. These are spatial loops though, whose properties are captured by our former analysis.}. Asymmetric configurations (for example with one boundary having a genus-g topology and the other being a sphere $S^2$), could potentially exist but are much harder to construct and study and we shall refrain from doing so here.

Considering loop cross-correlators instead, it is possible to find a worldsheet surface that ends on two wound loops around two cycles $\mathcal{C}_{1,2}$ on the two boundaries each having the topology of a Riemann surface $g$. This is possible since the wormhole does not have compressible bulk cycles and hence it can support a worldsheet surface that does not close off anywhere in the bulk that just extends between the two boundary cycles $\mathcal{C}_{1,2}$. This is in contrast with the single Wilson loop case where a disk worldsheet surface does not even exist on the wormhole background. This also means that there is no competition between a two-loop factorised and a non factorised configuration. We can then draw the following interpretations/conclusions of this result (in fact these are generalisable and hold in any number of dimensions):

\begin{itemize}

\item {\bf Thermal/Polyakov loops:} In the case that each wormhole boundary exhibits a thermal cycle $S^1_\beta$, the only bulk worldsheet configuration that ends on the two Polyakov loops, is that of a single connected surface. This means that while each individual $\langle W_P(\mathcal{C}_i )\rangle$ vanish, the connected correlator $\langle W_P(\mathcal{C}_1 ) W_P(\mathcal{C}_2 ) \rangle$ does not\footnote{Exept of geometric considerations, this statement also depends on the presence or not of zero modes for the bulk $B$-field, when considering a bulk worldsheet with the topology of a two boundary cylinder. More details are provided at the end of this section.}. This leads to some form of unbroken center symmetry due to $\langle W_P(\mathcal{C}) \rangle_{1,2} = 0$ on the wormhole background (a bulk surface does not even exist like in the case of AdS), indicating confining type of physics. On the other hand, there should also exist some notion of center symmetry breaking since for the pair of Polyakov loops\footnote{Insertions of a number of loops that is a multiple of $N$ \cite{Aharony:1998qu}, are non-zero even with unbroken center symmetry. In the bulk this manifests by a holographic construction of a baryon vertex (wrapped $NS5$ brane) on which strings can end~\cite{Aharony:1998qu,Gross:1998gk}.} $\langle W_P(\mathcal{C})_{1} W_P(\mathcal{C})_{2}  \rangle \neq 0$.

At first these statements might seem mutually exclusive. In fact they can be mutually consistent if the wormhole background breaks the boundary system's center symmetry to its diagonal part $Z^{(1)}_N \times Z^{(2)}_N \rightarrow Z^{diag.}_N$, and similarly the product
gauge symmetry from $SU(N)_{(1)} \times SU(N)_{(2)}$ for two disconnected geometries to $SU(N)_{diag.}$. In such a case, the ``axial/relative" center symmetry is (spontaneously or explictly) broken by the wormhole background and the observable that is sensitive to this broken axial part is precisely the loop cross-correlator that acquires a non-zero value. This means that while the usual one point Wilson loops (and multiple loops placed on a common boundary) are sensitive to the remaining unbroken diagonal center symmetry, there is still a mixed-observable (loop cross-correlator) whose expectation value indicates the broken axial part. This is a manifestation of what was termed \emph{a cross confining behaviour} in~\cite{BKP1}, where in holographic duals of Euclidean wormhole saddles the two a-priori individual boundary gauge groups need to fuse in the IR to their diagonal part. It also indicates that one has to introduce some appropriate form of interaction, that is very mild in the UV and very strong in the IR, in order to obtain results for the correlators that are consistent with having holographic dual Euclidean wormhole backgrounds/saddles.

Naively, this conclusion might also seem to cause a clash with the results and statements of~\cite{Schlenker:2022dyo}, since as we argued before the Euclidean wormhole background plays a role only when considering observables affected by high energy above the threshold (deconfined) states. The resolution of this apparent paradox lies in understanding that while the total combined $QFT$ system is always in a global singlet state (due to the unbroken diagonal center/gauge symmetry) indicating a confining behaviour, its subsystems are still allowed to transform in non trivial representations of their gauge group $\mathcal{G}_{1,2}$ that should nevertheless be tightly cross-correlated. 
In other words non-trivial representations and the relaxation of the Gauss' law constraint for each individual subsystem appears only at high energies above the BH threshold\footnote{A simple example was given in \cite{BKP2} in terms of representation correlated copies of matrix quantum mechanics. We analyze this example further in the next paragraph.}.
The bulk manifestation of this interesting behaviour is exactly the presence of the dual wormhole saddle. 
We should emphasize once more at this point that while the center symmetry $Z_N$ of a single gauge theory cannot be broken at finite volume, the symmetry breaking pattern $Z^{(1)}_N \times Z^{(2)}_N \rightarrow Z^{diag.}_N$ to the diagonal part that we advocate here, can in principle manifest itself even at finite spatial volume, since one can form a singlet by taking appropriate tensor products of representations $R_1 \otimes R_2$, where $R_{1,2}$ belong to each individual copy and this is true even for systems with finite degrees of freedom, such as simple spin systems. Once more, the only physical requirement and constraint is preserving Gauss' law for the entire combined system and not for the individual subsystems. 

The simplest example of a confining theory that exhibits an analogous symmetry breaking pattern, is 2d Yang-Mills (2d YM) on the cylinder \cite{Gross:1994ub}, coupled to one dimensional matrix quantum mechanics (MQM) living on the two $S^1$ boundaries of the cylinder~\cite{BKP2}, parametrised by $z \in [-L,L], \, \tau \in [0,\beta) \sim \tau + \beta$. The partition function of 2d YM on the cylinder is known to take the form~\cite{Gross:1994ub}
\be\label{cylinder2dYM}
Z_{cyl.}^{2d-YM}(U_1, U_2) = \sum_R \chi_R(U_1) \chi_R(U_2^\dagger) e^{- \frac{A g^2_{YM}}{N} C^{(2)}_R} \, ,
\ee
with $R$ denoting the $U(N)$ (or $SU(N)$) representations and $U_{1,2}$ the two holonomies of the gauge field around the $S^1$'s at the ends of the cylinder with area $A$ and coupling $g_{YM}$.
This cylinder amplitude is invariant only under the combined diagonal center transformation of the two Wilson loops/characters $\chi_R(U) \rightarrow \chi_R(g_c \, U) = g_c^{|R|} \chi_R(U)$, since the two characters that appear in \eqref{cylinder2dYM} transform in an opposite way. We can then imagine that we couple this cylinder amplitude to two separate one-dimensional gauged (adjoint) Matrix Quantum Mechanics living at the ends of the cylinder using a minimal gauging procedure and identifying the two MQM gauge fields with the value of the two-dimensional gauge field on the two boundaries of the cylinder (i.e. $A_\tau(z=\pm L , \tau) = A^{(1,2)}_\tau(\tau)$) \cite{BKP2}. The boundary degrees of freedom were chosen to preserve center symmetry in their action, so that it is not completely and explicitly broken by construction. The partition function of the combined system is \cite{BKP2}
\bea\label{pfsystem1}
Z_{system} &=& \sum_{R} \int D U_1 \int DU_2 \, \chi_R (U_1) Z^{MQM}_1 (U_1)  \chi_R (U_2^\dagger) Z^{MQM}_2 (U_2^\dagger) e^{-  {A g_{YM}^2}  C_R^{(2)} }  \, \nn \\
&=&  \sum_{R: \, (c.s.)}  e^{-  {A g_{YM}^2}  C_R^{(2)}} Z_{R}^1(\beta) Z_{\overline{R}}^2(\beta)  \, .
\eea
Due to the preservation of center symmetry by the boundary MQM degrees of freedom, the only representations that appear in the resulting (diagonal) final sum, are center symmetric representations (c.s.).
So long as we keep $A_{YM} = A g_{YM}^2$ finite, this also means that if we insert in eqn. \eqref{pfsystem1} single probe Polyakov loops
of representations that are non-center symmetric (such as the fundamental $\tr U$), then their expectation value will be zero. On the other hand for pairs of probe Wilson loops transforming in conjugate representations (even if non-center symmetric), the resulting expectation value is in general non zero. This behaviour is therefore the same as the one we found for the two-point cross correlator in the wormhole case.


\item {\bf Spatial loops:} A spatial loop (that is independent of the thermal $S^1_\beta$), behaves basically as in the case analysed at the beginning of section~\ref{WLprobes}. For a single spatial loop one finds an area law behaviour for large relative loop size $\langle W(\mathcal{C})  \rangle \sim e^{- A(D_\mathcal{C}^{bound.})}$. The general argument is as follows: As we described in section~\ref{WLandHP}, the expectation value of the loop always scales with the bulk area of the worldsheet dual surface (disk) $\langle W(\mathcal{C})  \rangle \sim e^{- A(D^{bulk})}\, , \, \mathcal{C} = \partial D^{bulk}$.  When the scale factor of the bulk geometry bounces (or has an inherent IR scale), the dominant worldsheet dual configuration for large loop size gets attracted to the IR and then the worldsheet disk area law effectively becomes a boundary area law $A(D^{bulk}) \simeq A(D^{bound.}_\mathcal{C})$.

If the cross-correlator factorises at large-N, its expectation value takes a product form of two disk contributions to leading order
\be
\langle W(\mathcal{C}_1)  W(\mathcal{C}_2)  \rangle \sim \langle W(\mathcal{C}_1)   \rangle \langle   W(\mathcal{C}_2)  \rangle \sim e^{- A(D^{bulk}_1) - A(D^{bulk}_2)} \, , \qquad \partial D^{bulk}_i = \mathcal{C}_i \, .
\ee
This holds up to a (connected) prefactor due to bulk perturbative mode exchanges between the two worldsheet disks. The analysis and physics then is similar to the single loop example and for large enough loops there is a boundary area law $A(D^{bulk}) \simeq A(D^{bound}_\mathcal{C})$ for each factor. On the other hand, in the case that the leading contribution to the cross-correlator is that of a connected surface between the two loops having the topology of a cylinder $S^1 \times I$, we find that
\be
\langle W(\mathcal{C}_1)  W(\mathcal{C}_2)  \rangle_c \sim e^{- A (Cyl. = S^1 \times I)} \, .
\ee
In this case the scaling is typically of a perimeter type, since the bulk cylinder area scales with the boundary $S^1$ loop length and the (regulated) distance between the two loops through the bulk is approximately a constant\footnote{This distance is infinite because the bulk surface reaches the asymptotic AdS boundaries and one has to regulate and renormalise it as in the case for a single loop.}. This result is also consistent with our previous arguments that the system exhibits a symmetry breaking pattern $SU(N)_1 \times SU(N)_2 \rightarrow SU(N)_{diag.}$, and while the single Wilson loops once more probe the remaining diagonal confined $SU(N)_{diag.}$, the loop cross-correlator is a probe of the broken (deconfined) axial part.

\end{itemize}

Let us finally emphasize that similarly to the case of the computation of holographic Wilson loops on geometries with a single boundary, one should also be careful about the presence and effects of the bulk $B$ field, since it could in principle render the expectation value of the loops or their correlators to zero. In the case of single Wilson loops, the effect of the $B$-field is exactly the same as in the single boundary geometries - for finite volume boundaries one has to integrate over the phase: $\Phi = \int_D B$, rendering the result equal to zero (of course as we argued before in some cases there is no bulk worldsheet surface ending on the loop to start with, so that the expectation value of the Wilson loop is already zero for stronger topological reasons). What is novel in the wormhole backgrounds is the effect of the $B$-field to the two point connected cross correlator with the topology of a cylinder. In particular one finds that one should then perform an integral over possible zero modes of the cross correlator phase given by
\be
e^{i \Phi_{cross}} = e^{i \int_{S^1 \times I} B} \, .
\ee
In the simplest case that the two boundaries are $S^d$ (and not $S^1_\beta \times S^{d-1}$ as in the case analysed in section~\ref{WLandHP}, where $\Phi$ behaves as a local scalar field on $S^{d-1}$), there is no such zero mode and hence the loop cross-correlator remains non-zero. On the other hand if the two loops are wound around two asymptotic $S^1_\beta$'s, then there is a zero mode for $\Phi_{cross}$ on the compact transverse space rendering this cross correlator equal to zero.

This concludes our analysis of probe Wilson loops on wormhole backgrounds with and without a thermal cycle $S^1_\beta$.

\subsection{Wormholes from backreacted Wilson loops and ``entangled representations"}\label{BackreactedWLandworms}

So far we analysed Wilson loops and their holographic dual worldsheet surfaces as probes of the (wormhole) geometry. In this section, we shall consider the opposite limit where the loops backreact strongly on the geometry and discuss the possibility of constructing wormhole saddles, precisely using such heavy Wilson loops. As we mentioned in section~\ref{BackreactedWL}, in order for the loop to be a very heavy operator, we need to consider more general loops that are traces of (large) non-trivial representations $R$ of the boundary gauge group $U(N)$ (or $SU(N)$). Motivated by the holographic results  for the probe loop correlators (``cross-confining'' picture) and the explicit two dimensional model on the cylinder (eqn.~\eqref{pfsystem1}), we are led to consider operators that contain
a strong representation theoretic correlation between the groups of two independent copies of $\mathcal{N}=4$ SYM. In general, the Schwinger functional of the combined system (together with the Wilson-loop insertions) will take a form similar to eqn. \eqref{pfsystem1}\footnote{Here and in the rest, the choice of whether to use the conjugate in the second subsystem, depends on the relative choice of orientation for the loops. For supersymmetric gauge theories, this choice is very important, since it can lead to a breaking of supersymmetry due to boundary conditions, leading to a confining system as argued in \cite{VanRaamsdonk:2021qgv}.}
\be\label{pfsystemworm}
Z_{system} = \sum_R \, e^{-w(R)} \, \langle W_R \rangle_1 \, \langle \overline{W_R} \rangle_2 \, ,
\ee
with $W_R = \tr_R P \exp{ (i \oint A)}$ (and its appropriate generalisations in the case of supersymmetric gauge theories such as $\mathcal{N}=4$ SYM). The “entangled sum” in \eqref{pfsystemworm}, can be thought of as arising either from integrating out some (topological degrees of freedom) as in \cite{BKP2}, or from a representation theoretic analogue of the thermofield double construction.
The usual thermofield double state 
\be
|TFD \rangle = \frac{1}{\sqrt{Z(\beta)}} \sum_{n} e^{- \frac{\beta}{2} E_n} |E_n \rangle_1 \otimes  |E_n \rangle_2 \, ,
\ee
correlates the energies of the two subsystems and creates a Lorentzian wormhole (the Einstein Rosen bridge).

In contrast, our ``representation theoretic entanglement" can be defined even in a purely Euclidean setting\footnote{In this case we do not need the presence of an $S^1$ factor in our manifold, that can be interpreted as Euclidean time upon analytic continuation. For example we can consider Wilson loops for theories living on $S^4$'s.} and exhibits some crucial differences with its energetic counterpart. For example, using Wilson loop operators an analogous $|R \, D \rangle$ state can be defined via the set of equations
\begin{align}\label{repanalogueTFD}
    \left| R\, D  \right\rangle &= \sum_R e^{-w(R)} \left| R \right\rangle_1 \otimes \left| R \right\rangle_2 \nonumber \\
\langle (P e^{ i \oint A})_1 ; (P e^{ i \oint A})_2   \left| {R\, D} \right\rangle  &= \sum_R e^{-w(R)} \, \left(  {W_R} \right)_1 \, \left(  \overline{W_R} \right)_2 \, .
\end{align}
If we demand that $|R\rangle$ label the different irreducible representations of $U(N)$, then they do form an orthogonal basis and hence one can recover some notion of density matrix \\ $\hat{\rho} = | R\, D \rangle \langle R\, D | \,$ and partial trace, but this only acts in the space of representations and does not capture the entire field theory path integral and Hilbert space of states. In other words, fixing a representation that can be viewed as a superselection sector, one still needs to sum over all energy eigenstates of the Hamiltonian in this sector\footnote{Some models where both summations appear and can be explicitly analysed, are models for the non singlet sectors of matrix quantum mechanics~\cite{BKP2,Betzios:2022pji}.}. In particular to transition from \eqref{repanalogueTFD} to \eqref{pfsystemworm}, one needs to further explicitly path integrate over all the fluctuations of all the fields in the two subsystems. The representations $|R\rangle$ then play a role analogous to $|\alpha \rangle$ states on which $Z_{system}$ factorises~\cite{Betzios:2022oef}.

Let us now try to understand a few more details of the generic physical properties of sums like \eqref{pfsystemworm} and in particular about the structure of the global and local (gauge) symmetries of the combined system and its subsystems (see also the comments in the introduction). Imagine we start from the decoupled QFT system that has a product symmetry $\mathcal{G}_1 \times \mathcal{G}_2$ (these can be either local or global symmetries). While each term in the sum \eqref{pfsystemworm} preserves a product subset of the original symmetry (for example two half-BPS Wilson loops in two copies of $\mathcal{N}=4$ in a fixed representation would preserve $16_1 \times 16_2$ supersymmetries of the original $32_1 \times 32_2$ of the two copies), the full sum
can exhibit saddles that either preserve or break this symmetry to its diagonal part. In particular a factorised geometric saddle would preserve an independent product symmetry, while a wormhole type of saddle inevitably breaks it to $\mathcal{G}_1 \times \mathcal{G}_2 \rightarrow \mathcal{G}^{sub.}_{diag.}$ with $ \mathcal{G}^{sub.}_{diag.} \subset \mathcal{G}_{diag.}$ a subset of the diagonal symmetry\footnote{Apart from the arguments we presented regarding gauge symmetries, in the case of global symmetries the common bulk gauge field and Gauss' law constraint, once more lead to a breaking of product global boundary symmetries to their diagonal part as described in the introduction.}. One can then further argue that the resulting backreacted (wormhole) geometry should have two asymptotic regions and be glued in the IR (since each loop operator average is a relevant deformation affecting the IR of the dual geometry). In the case of high enough supersymmetry (i.e. 1/2-BPS sector of $\mathcal{N}=4$ SYM), it is possible to analyse the system~\eqref{pfsystemworm} in great detail both from the bulk and the QFT side, using supersymmetric localization techniques for 1/2-BPS Wilson loops, that reduce the functional integral to a coupled matrix integral \cite{JiHoon}. 

\section{Discussion}\label{Discussion}

In this work we emphasised the importance of analysing Wilson loop observables in holographic gauge theories in the context of single and multi-boundary dual bulk manifolds. Apart from refining the construction by~\cite{Schlenker:2022dyo}, they can in general be used as probes of the effective bulk geometry at strong coupling indicating the transition between confining and de-confining behaviour. In the case of a non-trivial thermal $S^1_\beta$ we have a further option - that of a Polyakov loop that is sensitive to center symmetry breaking. The analysis of such observables on Euclidean wormhole backgrounds in section~\ref{WormholesandWL} led to formulating a picture of wormhole saddles as possibly being formed in systems of large-N gauge theories that exhibit the symmetry breaking pattern $SU(N)_1 \times SU(N)_2 \rightarrow SU(N)_{diag.}$ (and similarly for any other product symmetries).

Moreover, in sections~\ref{BackreactedWL} and~\ref{WormholesandWL} we described how such observables themselves, when being appropriately ``heavy", can give rise to dual backreacted geometries either with a single or multiple boundaries and clarified the connection with our previous work~\cite{BKP1,BKP2}. In particular the resulting entangled sums of the type given in eqns.~\eqref{correlatedloops} and~\eqref{pfsystemworm} can also be thought of as Euclidean counterparts of the thermofield double construction, see section~\ref{BackreactedWLandworms} and~\cite{Betzios:2022oef}. Throughout this work, we tried to be as general as possible. In explicit examples further details can be given and more refined physics are expected to be uncovered. We shall now close with a few more specific comments for future directions.

\paragraph{Loop operators across dimensions and the AGT correspondence} It is known that there exist deep relations between four-dimensional $\mathcal{N}=2$ four dimensional gauge theories and Liouville theory on Riemann surfaces under the name of the AGT correspondence~\cite{Alday:2009aq}. This correspondence, carries over also for loop operator observables~\cite{Drukker:2009id,Alday:2009fs}. In particular the relation can be summarised by
\be
\textit{Loop operator in} \, \, \mathcal{T}_{g,n} \quad \leftrightarrow \quad\textit{Liouville loop operator in} \, \, \mathcal{C}_{g,n} \, ,
\ee
where $\mathcal{T}_{g,n}$ symbolises the particular $\mathcal{N}=2$ gauge theory and $\mathcal{C}_{g,n}$  symbolises the Teichmuller space (the
universal covering space of the moduli space $\mathcal{M}_{g,n}$ of complex structures of punctured Riemann surfaces). This then means that perhaps one can study loop operator observables in Liouville theory and deduce from them properties of wormholes in $\mathcal{N}=2$ gauge theories\footnote{One would need to analyse a version of the AGT correspondence for $SU(N)$ gauge theories at large-N.}.

\paragraph{Algebra of Wilson loop observables? } It would be interesting to understand our analysis, from the recent perspective on the algebraic properties (von Neumann algebras) of observables (see~\cite{Witten:2023xze} and references within). In particular the HP transition (at finite transverse volume) signals a distinction between a fundamental type-I algebra and an emergent type-III von Neumann algebra for local operators at large N and infinite coupling~\cite{Engelhardt:2023xer}. As we know though, gauge theories have more physical structure than that accessible by purely local operators. To this end we are not aware of any analogous general classification of the algebraic structure of sets of non-local operators such as Wilson loops. This seems to be an outstanding mathematical problem with important physical ramifications, since the expectation value of a Polyakov loop is able to distinguish between the two phases of the HP transition (order parameter) and hence could capture even the transition between different types of von Neumann algebras for local operators. For holographic duals to spacetime wormholes, this analysis becomes even more intricate as we found in section~\ref{BackreactedWLandworms}, since it is possible to construct a representation theoretic analogue of the thermofield double, that seems to have an even more clear interpretation in a purely algebraic operator setting. This algebraic analysis for loop operators seems to be more concrete and tractable in the supersymmetric setting of $\mathcal{N}=2$ gauge theories via the AGT correspondence as alluded to above. We hope to report on this in the future.

\section*{Acknowledgements}\label{ACKNOWL}

We wish to thank Ofer Aharony, Costas Bachas, Yiming Chen, Nava Gaddam, Kristan Jensen, Arjun Kar, Elias Kiritsis, Ji-Hoon Lee, Juan Maldacena, Kyriakos Papadodimas, Mark van Raamsdonk, Gordon Semenoff and Ioannis Tsiares for various discussions on topics related to this work. We also wish to thank the participants of the \href{https://danninos.wixsite.com/psi-conf/psi2023}{Physics Sessions Initiative}, 
for their comments on our ideas presented there and the anonymous referee for suggestions that helped us to clarify and improve the text. 

\noindent The research of P.B. is supported in part by the Natural Sciences and Engineering Research Council of Canada. Research at Perimeter Institute is supported in part by the Government of Canada through the Department of Innovation, Science and Economic Development and by the Province of Ontario through the Ministry of Colleges and Universities. P.B. and O.P. acknowledge support by the Simons foundation.

\appendix
\section{The pinching limit for a genus two boundary surface and the dual CFT}\label{Pinchinggenustwo}

In this appendix we consider briefly the pinching limit of three dimensional holographic hyperbolic manifolds $\mathcal{M} \equiv X$ with a boundary $\partial  \mathcal{M} \equiv \Sigma$  being a genus two Riemann surface. One can define a renormalised volume for such manifolds in several different ways as reviewed in~\cite{Schlenker:2022dyo}. Some of them are more intrinsically geometric and abstract, but the simplest such definition from a physicist's point of view is in terms of the volume functional 
\be
\frac{V_X}{4 \pi G_N} = \frac{1}{4 \pi G_N} \int_X d^3 x \sqrt{g} - \frac{1}{8 \pi G_N} \int_\Sigma d^2 x \sqrt{h} K \, ,
\ee
where $V_X$ is the volume of $X$ and $K$ is a boundary Gibbons-Hawking (GH) term (if the manifold has a boundary). For on-shell solutions of pure three dimensional gravity this functional coincides with the Einstein-Hilbert functional (we set the $AdS$ scale to one)
\be
S = - \frac{1}{16 \pi G_N} \int_X d^3 x \sqrt{g} \left(R + 2 \right) - \frac{1}{8 \pi G_N} \int_\Sigma d^2 x \sqrt{h} K
\ee
evaluated on-shell (upon using Einstein's equations).
These quantities are divergent and need regularization and renormalization via the use of appropriate local counterterms on the conformal boundary $\Sigma$. There is a very well developed procedure that can achieve this (Holographic renormalization), see~\cite{Henningson:1998gx} and~\cite{Schlenker:2022dyo} for more details.

After renormalizing the volume of the manifold in some scheme (we will call this renormalized volume $V_{R(X)}$), one finds that it can still diverge when considering a pinching limit of a (bulk) contractible curve $\gamma$, in a precise fashion (see~\cite{Schlenker:2022dyo})
\begin{equation}
V_{R(X)} \leq V_{C(X)} - \frac{1}{4} L_h(l) + const. \, .    
\end{equation}
In this formula $C(X)$ is called the convex core of $X$
(it is a finite quantity not affected by the pinching procedure) and $L_h(l)$ is called the length of the measured lamination of the convex core. Physically it corresponds to a particular limiting procedure on the length of an appropriate set of curves $l$ (lamination) computed with the induced metric on the boundary $h$.
This is the quantity that diverges in the pinching limit (the length $L_h(l)$ depends on the size $b$ of the boundary pinching cycle/curve $\gamma$ in a non-trivial fashion). For a detailed analysis of these notions and proofs of various theorems the reader can consult the extended appendices in the last version of~\cite{Schlenker:2022dyo}.

Let us now assume a dual two dimensional CFT with Virasoro symmetry. The dual CFT partition function on a genus two Riemann surface takes the form  (in the so called "dumbbell" channel) \cite{Collier:2019weq}
\begin{equation}\label{dumbbellg2}
    Z_{g=2} = \sum_{\mathcal{O}_1} \sum_{\mathcal{O}_2} \sum_{\mathcal{O}_3}  C_{\mathcal{O}_1  \mathcal{O}_1  \mathcal{O}_3  } C_{\mathcal{O}_2 \mathcal{O}_2 \mathcal{O}_3 } \, \vert \mathcal{F}_{g=2}(\Delta_1, \Delta_2, \Delta_3)  \vert^2 \, ,
     \end{equation}
where $\Delta_i$ label the conformal dimensions of the operators and with $\mathcal{F}_{g=2}$ label the appropriate conformal blocks at genus two. In the relevant pinching limit, that separates the genus two Riemann surface in two tori, the genus two conformal blocks simplify into the product of two one-point torus conformal blocks, with independent moduli $\tau_{1,2}$
\begin{equation}
 \mathcal{F}_{g=2}(\Delta_1, \Delta_2, \Delta_3) \rightarrow b^{\Delta_3 - \frac{c}{24}}  \mathcal{F}_{g=1}(\Delta_1 ; \Delta_3 ; \tau_1) \mathcal{F}_{g=1}(\Delta_2 ; \Delta_3 ; \tau_2)  \, ,
\end{equation}
where $\Delta_3$ is the conformal dimension of the external one-point in the pinching limit and $b\rightarrow 0$ parametrises the pinching cycle/curve $\gamma$.  The partition function in the pinching limit then takes the form
\begin{equation}\label{pinchedPF}
 Z_{pinch} = \sum_{\mathcal{O}_1} \sum_{\mathcal{O}_2} \sum_{\mathcal{O}_3}  C_{\mathcal{O}_1  \mathcal{O}_1  \mathcal{O}_3  } C_{\mathcal{O}_2 \mathcal{O}_2 \mathcal{O}_3 } \, b^{2 \Delta_3 - \frac{c}{12}} \, \vert \mathcal{F}_{g=1}(\Delta_1 ; \Delta_3 ; \tau_1) \mathcal{F}_{g=1}(\Delta_2 ; \Delta_3 ; \tau_2)   \vert^2 \, ,   
\end{equation}
with the most relevant $\mathcal{O}_3$ operator being the identity. The result for all the sub-threshold operators is therefore diverging when $b \rightarrow 0$. The same type of divergence appears also in the $G \rightarrow 0$ limit of the gravitational path integral in which the bulk partition function is approximated by the renormalised on-shell action (volume)
\be
Z_{bulk} \sim e^{- V_R(X)/4 \pi G} \, ,
\ee
which diverges for $ V_R(X) \rightarrow - \infty$. On the other hand, for all the operators above the black hole threshold, the pinching limit is smooth (and in fact $Z_{pinch}({\Delta_3 > c/24} ) \rightarrow 0$, reflecting the fact that in three dimensional pure gravity there is no $I \times T^2$ wormhole saddle with $T^2$ asymptotic boundaries) - the related Cotler-Jensen $I \times T^2$ wormhole amplitude~\cite{Cotler:2020ugk} is an $O(1)$ contribution to the path integral.

We therefore observe that the gravitational calculations are consistent with the CFT pinching formula~\eqref{pinchedPF}, for the subthreshold states of holographic CFTs, that can be identified with hyperbolic manifolds that factorise in the pinching limit. It is an interesting future exercise to consider a similar pinching procedure in the presence of heavy operators placed on the genus two Riemann surface, either local or non-local (such as Verlinde loops). In this case we expect that the heavy operators backreact on the dual geometry resulting to an actual wormhole saddle with $T^2$ boundaries.




\end{document}